\newcommand{\abinitio}{\textit{ab-initio }}

\newcommand{\AAA}{$\textnormal{\AA}$ }

\documentclass[aps,prl,reprint, groupedaddress,longbibliography]{revtex4-1}
\usepackage{graphicx}
\usepackage{amsmath}
\usepackage[caption=false]{subfig}
\usepackage{float}
\usepackage{amssymb}
\usepackage{amsmath}

\begin{document}


\title{Electrical-equivalent van der Waals gap for 2D bilayers}


\author{Dhirendra Vaidya}
\email[]{dhirendra22121987@gmail.com}
\affiliation{Department of Electrical Engineering, IIT Bombay, Mumbai 400076, India}
\author{Saurabh Lodha}
\affiliation{Department of Electrical Engineering, IIT Bombay, Mumbai 400076, India}
\author{Swaroop Ganguly}
\email[]{swaroop.ganguly@gmail.com}
\affiliation{Department of Electrical Engineering, IIT Bombay, Mumbai 400076, India}


\date{\today}

\begin{abstract}
 Vertical stacks of two-dimensional (2D) materials, separated by the van der Waals gap and held together by the van der Waals forces, are immensely promising for a plethora of nanotechnological applications. Charge control in these stacks may be modeled using either a simple electrostatics approach or a detailed atomistic one. In this paper, we compare these approaches for a gated 2D transition metal dichalcogenide bilayer and show that recently reported electrostatics-based models of this system give large errors in band energy compared to atomistic (Density Functional Theory) simulations. These errors are due to the tails of the ionic potentials that reduce the electrical-equivalent van der Waals gap between the 2D layers, and can be corrected by using the reduced gap in the electrostatic model. For a physical van der Waals gap (defined as the chalcogen to chalcogen distance) of 3 \AAA in a 2D bilayer, the electrical-equivalent gap is less than 1 \AA. For the example of band-to-band tunneling based ultra low-power transistors, this is seen to lead to errors of several hundred millivolts and more in the threshold voltage estimated from electrostatics.
\end{abstract}


\maketitle

Recently there has been burgeoning interest in 2D vertical stacks for a wide variety of electronic applications. When a 2D monolayer (ML) of transition metal dichalcogenides (TMD) or graphene is placed on top of another 2D ML, such a bilayer (BL) is held together by the weak van der Waals force between the two MLs. These homogeneous or heterogeneous BLs or multilayers may be called van der Waals (vdW) stacks; they have been imagined as the building blocks of many novel electronic \cite{Tania_Roy_ACSNANO_2014,Tania_Roy_ACSMI_2015,Tania_Roy_APL_2016,Deblina_Sarkar_Nature_2015, Xiao_Peng_small_2017_37MV_SS}, optoelectronic \cite{Fang_PNAS}, spintronic \cite{Dankert2017} and power devices \cite{Choong_Hee_Lee_APL_2017}. 

Possibly the most significant use of these 2D BLs would be in next-generation tunnel-FETs (TFETs), to fulfil the extreme energy-efficiency requirement for internet of things (IoT) applications projected by the International Roadmap for Devices and Systems  \cite{IRDS2016}. The workhorse of conventional electronics, viz. the CMOS switch, works by controlling the height of a thermal barrier; and therefore it needs at least $(kT/q)\textnormal{ln}10=60$ mV to  change the current by an order of magnitude; in other words, it has a subthreshold swing (SS) greater than 60 mV/decade. TFETs, on the other hand, work by controlling the thickness of a tunneling barrier; so they can, in principle, have an SS $<$ 60 mV/decade\cite{TFET_review} and are therefore of immense interest for extreme low-energy IoT applications. TFETs have been fabricated and studied using group IV \cite{Pandey_Datta_IEDM_2016, Choi_Liu_EDL_2007_Si_TFET, Tejas_Krishnamohan_IEDM_2008} as well as III-V \cite{Zhou_Xiang_III_V_TFET} semiconductors. However, because of the usually high density of interface states \cite{Pala_Esseni_TED_2013_TFET_Traps} - leading to trap-assisted tunneling, achieving an SS $<$ 60 mV/decade in these systems has proven to be difficult. This issue might be alleviated by using a vertically stacked 2D BL and controlling the tunnel current between the top and bottom MLs through gate voltages, leading to super-steep sub-threshold characteristics. Even though realizing a vertical 2D stack \cite{Yongji_Gong_Nature_mat_2014} of two MLs is a challenging task, several novel device ideas have been explored in some depth in the literature \cite{Mingda_Li_JAP_2014, Mingda_Li_JEDS_2015, J_Cao_TED_2016, Campbell_ACSNano_2015, Campbell_EDL_2017}. 

As in most electronic devices, electrostatics plays the leading role in the functionality of 2D devices, including BL devices. In prior work, e.g. \cite{Mingda_Li_JAP_2014, Mingda_Li_JEDS_2015, J_Cao_TED_2016, Campbell_ACSNano_2015, Campbell_EDL_2017}, the band alignments of the vertical 2D stacks have been modeled using simple electrostatics -  i.e. assuming charge sheets corresponding to the MLs, separated by the their physical distance. This ignores the spatial distribution of the ionic potentials, in particular their tails falling off from the atomic sites, and thereby underestimates (overestimates) the effective electrical thickness (separation) of the MLs. We will show that this can lead to significant errors in modeling the gate control when the two layers are separated by a distance comparable to the vdW gap (defined as chalcogen to chalcogen distance) viz. $\approx$ 3-3.5$\textnormal{ \AAA}$ \cite{Lu_Ning_Nanoscale_2014, Wang2017, Hu2016}.

\begin{figure}[t]
\centering
   \includegraphics[width=0.48\textwidth]{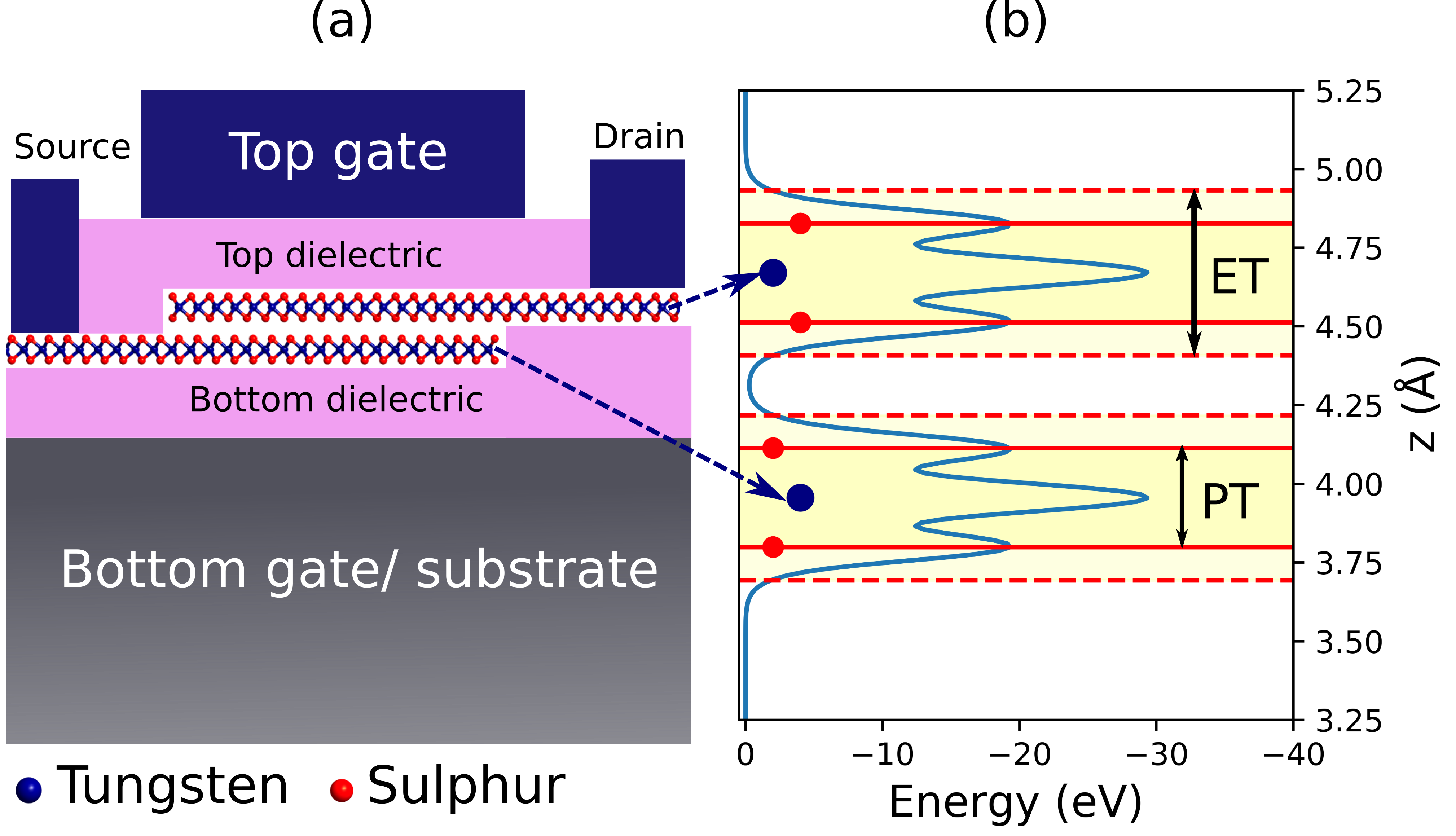}
   \caption{(a) WS\textsubscript{2} BL vdW TFET. In general the top and bottom MLs can be any suitable combination, however for the simplicity of the study we have considered 2D homojunction (BL) vdW TFETs. (b) Laterally averaged electrostatic potential energy of WS\textsubscript{2} BL vdW TFET. PT = physical thickness. ET = electrical thickness. PT is defined as the distance between the chalcogen atoms of same ML. ET is introduced to precisely account for the tails of the ionic potentials which extend beyond the lattice sites.} \label{fig:vdW_TFET}
\end{figure}

The aforementioned effect may be illustrated by considering the cross-section of a vdW BL and the ionic potentials therein. Figure \ref{fig:vdW_TFET}a shows the specific example of a homojunction WS\textsubscript{2} BL TFET. A vdW BL can in general be realized with vertical stacking of any suitable combination of 2D MLs. We choose a homojunction BL here for simplicity of illustration and intuitive explanation. Figure \ref{fig:vdW_TFET}b shows the laterally averaged electrostatic potential energy of WS\textsubscript{2} BL system in equilibrium. We can clearly see that the electrical thickness of each ML is larger than the physical thickness (defined by the lattice sites of the boundary chalcogen atoms) on account of the tails of the ionic potentials. A comparison of atomistic simulations using density functional theory (DFT) versus electrostatic simulations reveals that large errors are possible in the regular electrostatic simulations which ignore the effect of the ionic potential tails.

\begin{figure}[t]
\centering
   \includegraphics[width=0.43\textwidth]{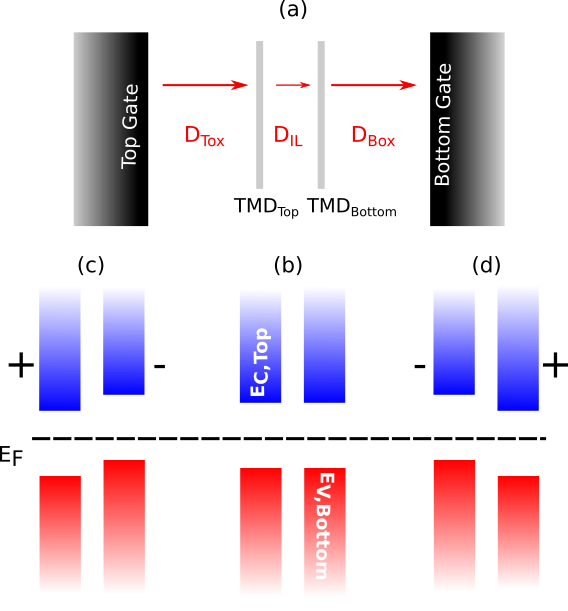}
   \caption{Illustration of the band alignment of the vdW stacks under the electric field. The polarity signs show the bias polarities on the top and bottom gates. (a) a gated TMD BL system. (b) shows the CB and VB of top and bottom MLs at zero bias. (c) and (d) shows the CB and VB of top and bottom MLs at non-zero bias.} \label{fig:band_alignment_illustration}
\end{figure}

We will begin with a discussion of the intuitive picture of vertical 2D MLs under an electric field. There we also explain the band alignment in these vertical 2D stacks, and present the popular purely electrostatic approach that has been used in modeling these vdW stacks. Following that, we discuss the tails of the ionic potentials and the necessity of including these for predictive modeling. Thereafter, we compare DFT and electrostatic simulations of the vdW stacks under an electric field, and propose a simple correction to the interlayer separation used in the electrostatic calculation in order to model the effect of the ionic potential tails. Lastly, the importance of the potential tail correction on device design is illustrated by comparing the threshold voltages for band to band tunneling (BTBT) obtained from the simple (uncorrected) electrostatic model versus our corrected one.

 Figure \ref{fig:vdW_TFET}a shows a four terminal electronic device. Such a device can be used as a $p-n$ junction diode\cite{Tania_Roy_ACSMI_2015, Tania_Roy_APL_2016}, a tunnel diode\cite{Campbell_ACSNano_2015} or a TFET\cite{Mingda_Li_JAP_2014,Mingda_Li_JEDS_2015}. This flexibility in application is possible because the top and bottom gate can modulate the band offsets of the top and bottom MLs.Figure \ref{fig:band_alignment_illustration}a shows a gated TMD bilayer system. CB and VB offsets in two MLs of the system can be modified by applying a gate bias. Figure \ref{fig:band_alignment_illustration}b shows the band alignment of the 2D MLs when no bias is applied. Under this condition, the band offsets are obviously zero as may be expected. Figure \ref{fig:band_alignment_illustration}c shows the band alignments when a positive voltage is applied to the top gate and a negative voltage is applied to the bottom gate. The positive potential on the top gate makes the top ML n-type and the conduction band minimum (CBM) is near the Fermi level; whereas the negative potential on the bottom gate makes the bottom ML p-type and the valence band maximum (VBM) moves closer to the Fermi level. The condition reverses on reversing the bias polarities (Figure \ref{fig:band_alignment_illustration}d). It is this gate control of band offsets in the vdW BL systems that opens up many interesting device application possibilities.

Since the scope of this paper is gate-controlled band alignment, we ignore the transport completely in what follows, and drop the source and drain terminals from our discussion (Figure \ref{fig:band_alignment_illustration}a).
The charge density ($\rho_{2D,T}$ and $\rho_{2D,B}$) in the 2D MLs ($\text{TMD}_{\text{Top}}$ and $\text{TMD}_{\text{Bottom}}$) can be related to the electric fields ($\mathcal{E}_{Tox}$, $\mathcal{E}_{IL}$ and $\mathcal{E}_{Box}$) using Gauss' law.

\begin{align}
D_{IL}-D_{Tox} &= \rho_{2D,T} \\
D_{Box}-D_{IL} &= \rho_{2D,B} \\
\epsilon_{IL}\mathcal{E}_{IL} - \epsilon_{Tox}\mathcal{E}_{Tox} &= \rho_{2D,T} \\
\epsilon_{Box}\mathcal{E}_{Box} - \epsilon_{IL}\mathcal{E}_{IL} &= \rho_{2D,B}
\end{align}
and

\begin{align}
\mathcal{E}_{Tox} &= \frac{(\chi_T+q\phi_{n,T})-(q\phi_{M,T}-qV_{TG})} {t_{Tox}} \\
\mathcal{E}_{IL} &= \frac{(\chi_B+q\phi_{n,B})-(\chi_T+q\phi_{n,T})} {t_{IL}} \\
\mathcal{E}_{Box} &= \frac{(q\phi_{M,B}-qV_{BG})-(\chi_B+q\phi_{n,B})} {t_{Box}}
\end{align}
where, $\chi_T, \chi_B$ are the electron affinities of top and bottom MLs. $t_{Tox}, \text{ } t_{IL} \text{ and } t_{Box}$ are the physical thicknesses of the top, bottom and inter layer dielectrics. $\phi_{M,T} \text{ and } \phi_{M,B}$ are the metal workfunctions of the top and bottom gates and $q\phi_{n,T} = E_{c,T}-E_F \text{ and } q\phi_{n,B} = E_{c,B}-E_F$. The electron and hole densities in each monolayer are then related to $\phi_{n,T} \text { and } \phi_{n,B}$ as,

\begin{align}
n_T &= \frac{g_{valley,T}m_{c,T}k_BT}{\pi \hbar^2}\text{ln} \left[\text{exp}\left(-\frac{q\phi_{n,T}}{k_BT}\right)+1\right] \\
n_B &= \frac{g_{valley,B}m_{c,B}k_BT}{\pi \hbar^2}\text{ln} \left[\text{exp}\left(-\frac{q\phi_{n,B}}{k_BT}\right)+1\right] \\
p_T &= \frac{g_{valley,T}m_{v,T}k_BT}{\pi \hbar^2}\text{ln} \left[\text{exp}\left(-\frac{q\phi_{p,T}}{k_BT}\right)+1\right] \\
p_B &= \frac{g_{valley,B}m_{v,B}k_BT}{\pi \hbar^2}\text{ln} \left[\text{exp}\left(-\frac{q\phi_{p,B}}{k_BT}\right)+1\right]
\end{align}
where, $q\phi_{p,T}=E_{g,T}-q\phi_{n,T}$ and $q\phi_{p,B}=E_{g,B}-q\phi_{n,B}$. $E_{g,T} \text{ and } E_{g,B}$ are the bandgaps of top and bottom 2D MLs respectively. The total charge density in each ML is given by,

\begin{align}
\rho_{2D,T} &= q(p_T - n_T + N_{D,T} - N_{A,T}) \\
\rho_{2D,B} &= q(p_B - n_B + N_{D,B} - N_{A,B})
\end{align}
We have solved equations (3), (4), (12) and (13) self-consistently, using a Newton-Raphson method.

 Figure \ref{fig:vdW_TFET}b shows the laterally averaged electrostatic potential of a WS\textsubscript{2} homojunction BL system when the interlayer separation is 4 \AA. It can be seen that the ionic potentials extend a few angstroms beyond the lattice sites of the chalcogens which are at the edges of the MLs. The extent of these tails into the gap between the MLs is comparable to the vdW gap ($\approx$ 3-3.5 \AA). 

\begin{figure}[t]
\centering
   \includegraphics[width=0.5\textwidth]{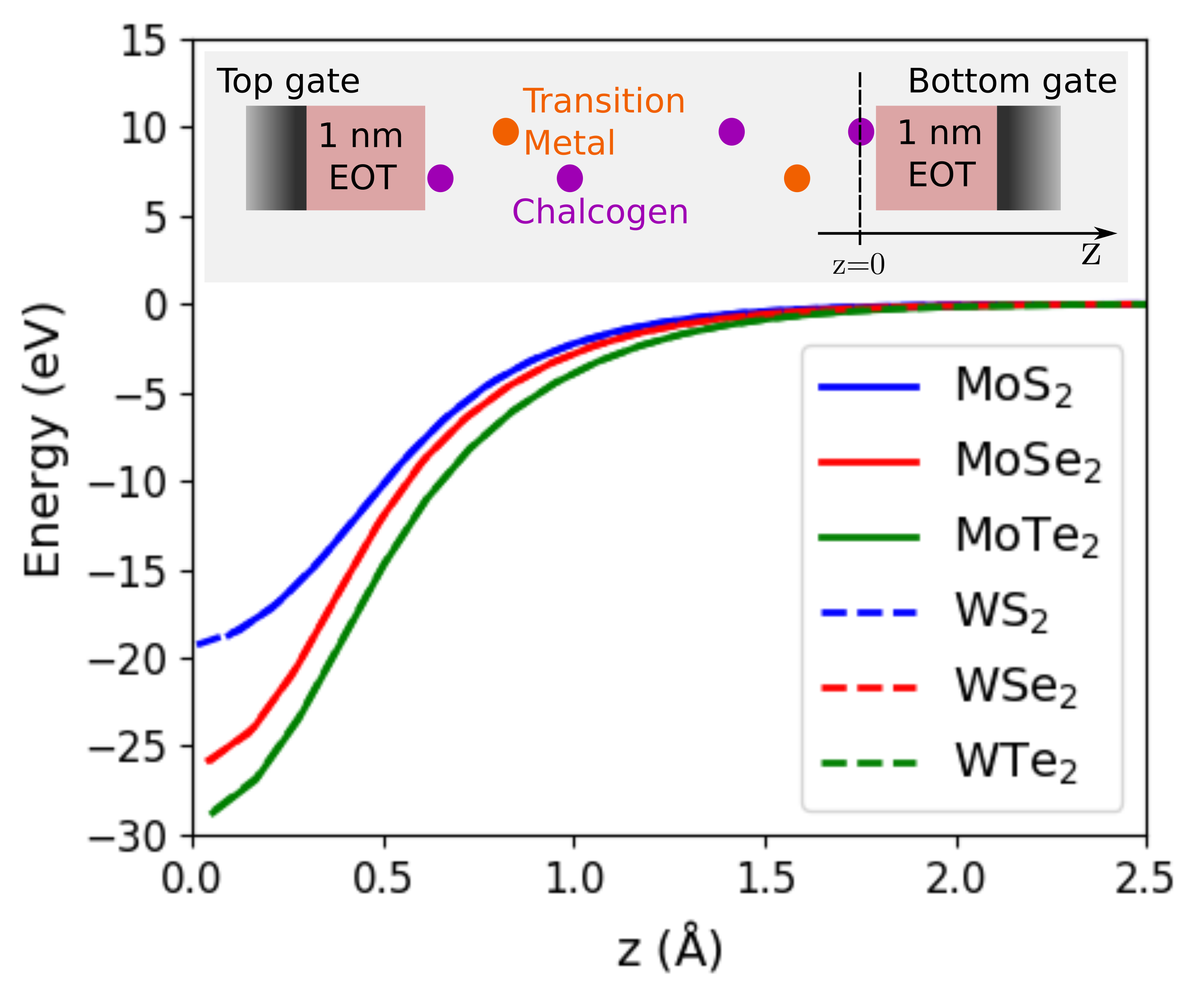}
   \caption{Laterally averaged electrostatic potential energy from boundary chalcogen lattice site into vacuum. As shown in the graphical inset, the lattice site of boundary chalcogens is at $z=0$. It can be seen that the laterally averaged electrostatic potential energy from the boundary chalcogen atoms of molybdenum dichalcogenides is nearly identical to that of the tungsten dichalcogenides, indicating the dependence of boundary chalcogen atoms on the potential tails.} \label{fig:Monolayer_tails}
\end{figure}

Assuming 2D charge sheets corresponding to the TMD MLs, separated by the physical distance between them, would seem to be a good approximation when the interlayer separation is large compared to the extent of the ionic potential tail. It is obviously suspect when these become comparable. However, this latter case is of great practical interest since the interlayer separation in vertical stacks of 2D MLs is expected to be the vdW gap. Figure \ref{fig:Monolayer_tails} shows the laterally averaged electrostatic potential energy, obtained from DFT simulations, for BL homojunctions of MoS\textsubscript{2}, MoSe\textsubscript{2}, MoTe\textsubscript{2}, WS\textsubscript{2}, WSe\textsubscript{2} and WTe\textsubscript{2} as a function of position, starting from the lattice site of the outermost chalcogens. Since the chalcogen atoms are on the edges of the TMD MLs, it is the tails of the chalcogen ion potentials that extend a few angstroms into the space between the two MLs. Now, in order to quantify the extent of the tail, we observe that the potential energy for MoS\textsubscript{2} and WS\textsubscript{2} decays to 10\% of the value that it has at the boundary chalcogen lattice site at a distance of 1.05 \AA. Thereafter we consider this 10\% value for the disulfides (viz. -1.93 eV) as the reference. We then observe that the potential energy for both MoSe\textsubscript{2} and WSe\textsubscript{2} reach that value at a distance of 1.12 \AAA from the boundary chalcogen atom; while for both MoTe\textsubscript{2} and WTe\textsubscript{2}, this distance is 1.25 \AA.

Before proceeding further, we discuss another possible explanation for the error in the pure electrostatic model that we considered: namely, the spread in the electron/hole wavefuctions. We modeled each 2D MLs as a delta-function potential well. Then the wavefunction of electrons and holes are of the form,

\begin{align}
\psi(z) &= \sqrt{\kappa}e^{\kappa (z-z_0)} \textnormal{ for } z<z_0\\
 & = \sqrt{\kappa}e^{-\kappa (z-z_0)} \textnormal{ for } z>z_0
\end{align}
where $z_0$ is the position of the ML. Now, the spatial distribution of electrons/holes can be captured by multiplying equations (8-11) by $|\psi(z)|^2$ from equations (14-15). This would also suggest a finite thickness of the MLs. However, using a 1D finite difference Poisson solver\cite{Vaidya2015} we found that the band alignments in vdW stacks are insensitive to the wave nature of electrons/holes for reasonable values of $\kappa$. Therefore we concluded that it is not the spread in the wavefunction that causes the error in the electrostatic model.

\section{Results and Discussion}

 A fair comparison between atomistic DFT and phenomenological electrostatic models requires, firstly, a consistent set of TMD ML parameters to be used therein. Thus, for the electrostatic simulations, we use parameters that are extracted from the DFT simulations; these are shown in Table \ref{Table:band_parameters}. The extracted values of the lattice constant, bandgap and effective masses are in reasonable agreement with previous abinitio studies and experiments\cite{MoS2_Latt_Const_1968, Chiu2015, Reshak2003}.

\begin {table}[t]
\caption {Band parameters of ML TMDs. $a_0$ is the lattice constant of TMD ML, $E_{g,K-K}$ is the bandgap of TMD ML determined at $K$-point, $\Delta E_{C,Q-K}$ ($\Delta E_{V,K-\Gamma}$) is the difference between the $Q$ ($\Gamma$) and $K$ valley minima (maxima) in conduction (valence) band.} \label{Table:band_parameters} 
\begin{center}
  \begin{tabular}{ ccccccc }
     \hline \hline
 TMD &$a_0$&$m_c$&$m_v$&$E_{g,K-K}$&$\Delta E_{C,Q-K}$&$\Delta E_{V,K-\Gamma}$\\
     &\AA&($m_0$)&($m_0$)&(eV)&(eV)&(eV)\\
 \hline
 MoS\textsubscript{2}& 3.132 & 0.581 & 0.629 & 1.910 & 0.022 & 0.192\\
 MoSe\textsubscript{2} & 3.247 & 0.619 & 0.678 & 1.658 & 0.030 & 0.399\\
 MoTe\textsubscript{2} & 3.496 & 0.603 & 0.717 & 1.166 & 0.120 & 0.532\\
 WS\textsubscript{2} & 3.131 & 0.334 & 0.432 & 2.010 & 0.113 & 0.125\\
 WSe\textsubscript{2} & 3.244 & 0.370 & 0.467 & 1.783 & -0.001 & 0.453\\
 WTe\textsubscript{2} & 3.498 & 0.366 & 0.474 & 1.226 & 0.061 & 0.6266\\
 \hline
  \end{tabular}
\end{center}
\label{Table:Band_Parameters}
\end{table}

\begin{figure}[t]
\centering
   \includegraphics[width=0.48\textwidth]{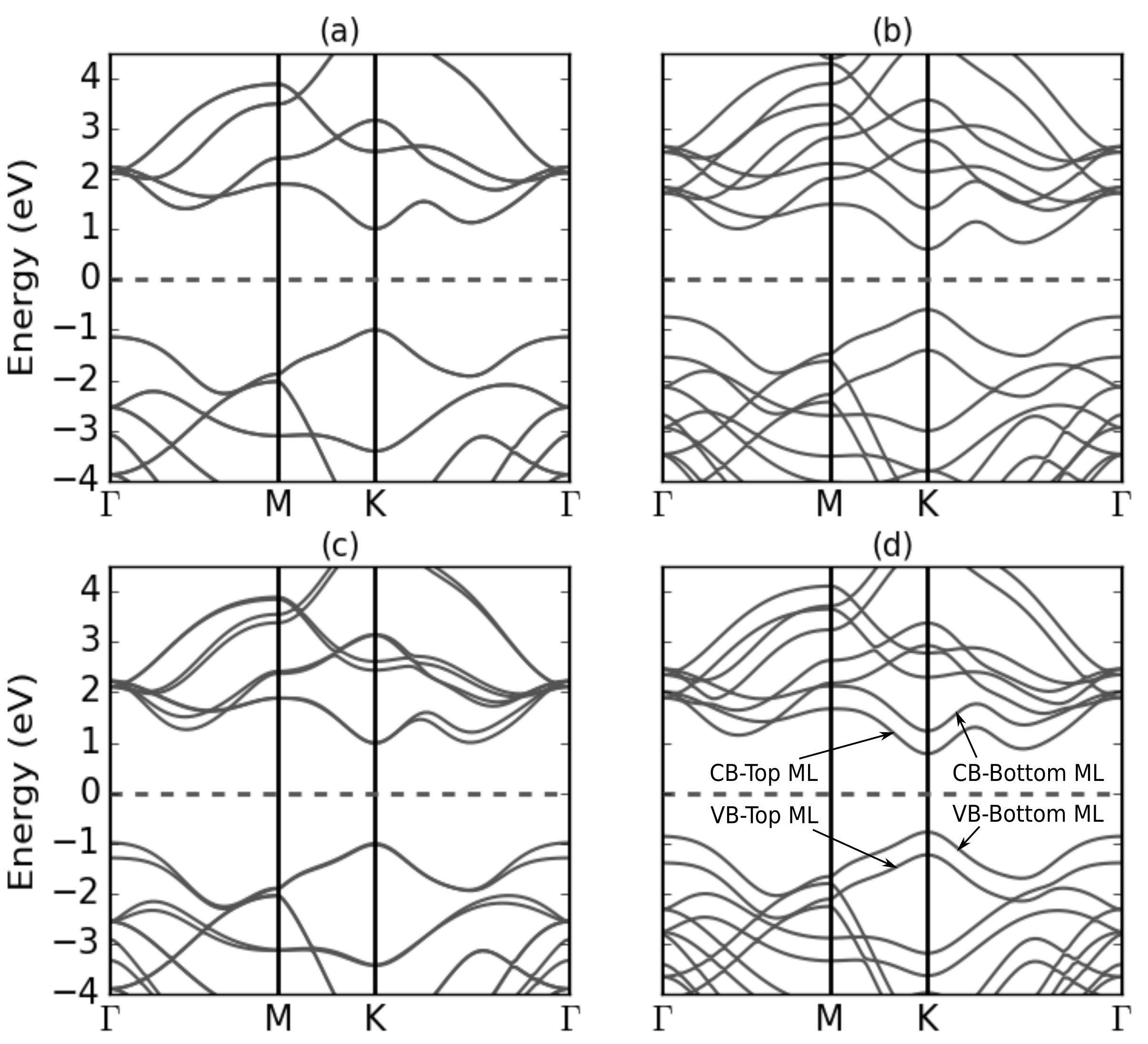}
   \caption{(a) Bandstructure of WS\textsubscript{2} BL system with 7 \AAA interlayer separation at zero gate bias. (b) Bandstructure of WS\textsubscript{2} BL system with 7 \AAA interlayer separation at a bias of $V_{TG} = +V$ and $V_{BG} = -V$. (c) Bandstructure of WS\textsubscript{2} BL system with 4 \AAA interlayer separation at zero gate bias. (d) Bandstructure of WS\textsubscript{2} BL system with 4 \AAA interlayer separation at a bias of $V_{TG} = +V$ and $V_{BG} = -V$.}
   \label{fig:bandstructures}
\end{figure}

\begin{figure*}[t]
\centering
   \includegraphics[width=1.0\textwidth]{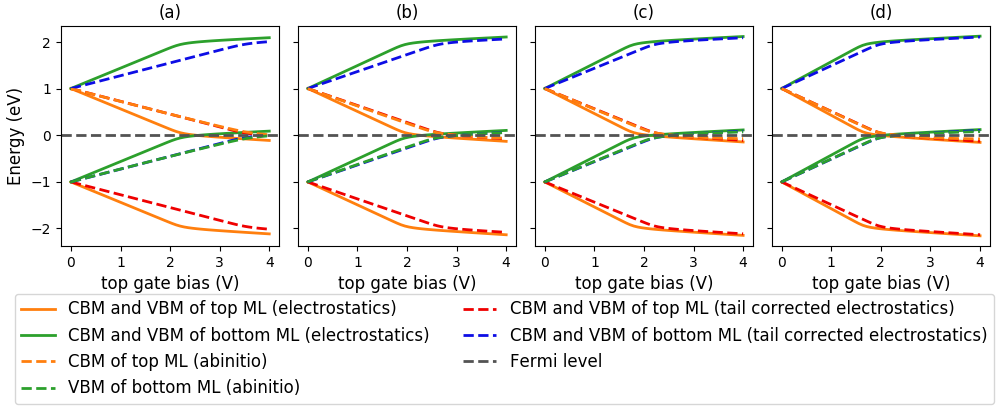}
   \caption{Comparison of electrostatic and DFT simulations of WS\textsubscript{2} BL system under the electric field. The top gate voltage ($V_{TG}$) is varied from 0 V to 4 V. The equal and opposite polarity bias applied to the bottom gate ($V_{BG}=-V_{TG}$). The interlayer separation in the BL system is (a) 4 \AAA, (b) 5 \AAA, (c) 6 \AAA and (d) 7 \AA. It can be seen that the electrostatic simulations agree with the DFT simulations if the the tail corrected interlayer separation is used, the two characteristics overlap. For WS\textsubscript{2} BL system this correction is 2.05 \AAA (each WS\textsubscript{2} ML contributes 1.025 \AA)} \label{fig:bands_comparison}
\end{figure*} 

\begin{figure}[t]
\centering
   \includegraphics[width=0.5\textwidth]{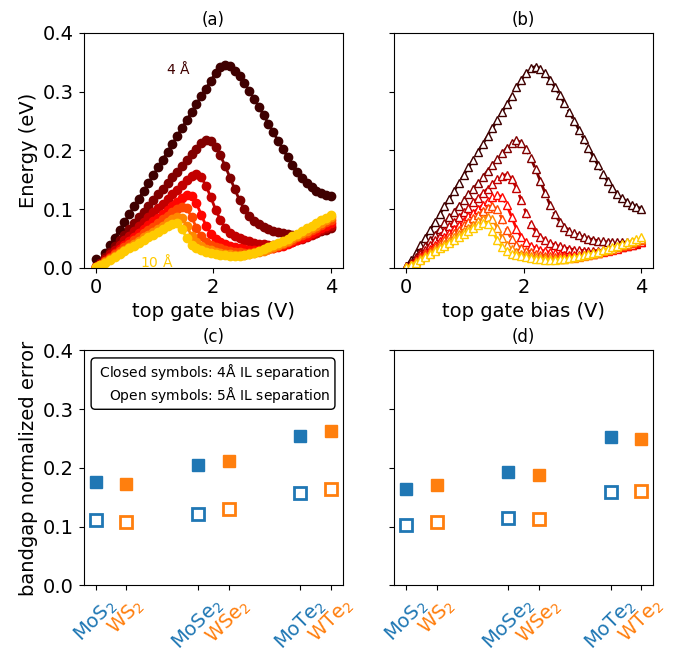}
   \caption{Error in electrostatic simulations (a) in CBM of top ML and (b) in VBM of bottom ML as a function of bias and interlayer (IL) separation. The interlayer separation is varied from 4 \AAA to 10 \AAA with step size of 1 \AA. It can be seen that the maximum error in CBM of top ML and VBM bottom ML is inversely proportional to the interlayer separation. (c) Bandgap normalized maximum error in CBM of top ML when interlayer separation is 4 \AA and 5 \AA, (d) Bandgap normalized maximum error in VBM of bottom ML when interlayer separation is 4 \AAA and 5 \AAA.}
   \label{fig:Error_Vs_bias}
\end{figure}

As depicted in figure \ref{fig:band_alignment_illustration}, the applied gate electric field modulates the band alignments in the two MLs. The band offsets can be inferred from the conduction band edges at $K$-point in the Brillouin zone. The bandstructure of the overall BL system such as the one shown in inset of figure \ref{fig:Monolayer_tails} is just the superposition of the ML bandstructures, which are degenerate at zero bias. This degeneracy is broken by an applied gate electric field. Then, one ML becomes n-type while the other ML becomes p-type (figure \ref{fig:band_alignment_illustration}). In the bandstructure of the BL system, the CBM can be seen shifted towards (away from) the Fermi level for the n-type (p-type) ML. The difference between the CBM of the two MLs gives the band offset ($\Delta E_C$). At higher biases, the CBM of the p-type ML moves far away from the Fermi level and mixes with the higher order bands of the other ML. This makes the direct measurement of the CBM of the p-type ML difficult but measuring the VBM of this ML at the $K$-point is straightforward. 

Figure \ref{fig:bandstructures} shows the bandstructures of BL systems of WS\textsubscript{2} having 7 \AAA (a and b) and 4 \AAA (c and d) interlayer separations. Figure \ref{fig:bandstructures}a is a bandstructure of a WS\textsubscript{2} BL system with interlayer separation of 7 \AAA under the zero-bias condition. Since the interlayer separation is 7 \AAA the two MLs in the BL system are practically isolated from each other. The bandstructure of the entire BL system is a doubly degenerate ML bandstructure. When the BL system is subjected to an electric field, the degeneracy breaks; this is illustrated in figure \ref{fig:bandstructures}b. Figure \ref{fig:bandstructures}b shows the bandstructure of the overall BL system for $V_{TG} = +V$ and $V_{BG} = -V$. Due to the applied gate field in this case, the top ML becomes n-type and the CBM of the top ML moves towards the Fermi level. The VBM of the bottom ML moves towards the Fermi level indicating that the bottom ML has become p-type. It can be seen in figure \ref{fig:bandstructures}b that the bandgap of the overall BL system at $K$-point is smaller than the ML bandgap. This is not true bandgap modulation with electric field as suggested by \citeauthor{Liu_Qihang_JPCC_2012}\cite{Liu_Qihang_JPCC_2012} but rather an effect of the potential drop across the BL system, with each ML retaining its ML bandgap value (as shown in the Supplementary Material). Figure \ref{fig:bandstructures}c (zero bias) and \ref{fig:bandstructures}d ($V_{TG} = +V$ and $V_{BG} = -V$) show the bandstructure of the overall BL system of WS\textsubscript{2} when the interlayer separation is 4 \AA. It can be seen that this is not exactly the double degenerate ML bandstructure at zeros bias. CB near $Q$-point (which is halfway between the $\Gamma$-point and $K$-point) and VB near $\Gamma$-point are not doubly degenerate. This is because of significant interlayer interaction at the smaller separation, which modifies the Hamiltonian of each layer. 

The interlayer interaction \cite{Liu_Nature_2014} of course increases as the separation is reduced. This necessitates modification of the BL Hamiltonian to incorporate off-diagonal interaction terms between the MLs, which would in turn modulate the solutions to the Schr{\"o}dinger equation, viz. the bandstructure. However, the methodology for solving the Poisson equation to capture electrostatic effects due to space charge would remain the same. To elucidate this methodology, while excluding the complication due to the interacting BL Hamiltonian, we constrain our discussion to interlayer separation of 4$\textnormal{ \AA}$ and more. However, we emphasize that the electrostatics treatment presented here would continue to hold for smaller separation.   

Figure \ref{fig:bands_comparison} shows the comparison between electrostatic and DFT simulations for a bilayer system of WS\textsubscript{2}. The comparison for other TMDs is presented in supplementary material. The top gate is biased from 0 V to 4 V and the opposite polarity bias is applied to the bottom gate. This makes the top ML n-type and the bottom ML p-type.  The DFT simulations are subject to the constraint of overall charge neutrality; because the total number of electrons are fixed by the number of atoms. For a reasonable comparison, the electrostatic simulations must also ensure overall charge neutrality. Applying opposite polarity bias to the top and bottom gates imposes overall charge neutrality in the BL system ($\rho_{2D,T}=-\rho_{2D,B}$). We compare DFT and electrostatic simulations for interlayer separation of 4, 5, 6 and 7 \AA. It can be seen that the electrostatic simulations and the DFT simulations are not in quantitative agreement. At a given top gate bias (+V) the electrostatic model suggests that the Fermi level is closer to the CB (VB) edge of the top (bottom) ML compared to the DFT simulations. The error shown in figure \ref{fig:Error_Vs_bias} shows that the electrostatic and atomistic simulations agree (disagree) for large (small) interlayer separation. Figure \ref{fig:Error_Vs_bias} shows that the error is maximum at the bias point where each ML is close to becoming degenerate n-type or p-type. Comparing the nearly intrinsic bias region ($V_{TG} = 0$ to $0.8$ V) to the extreme degenerate region ($V_{TG} > 3.0$ V), we see that the error is smaller in the former. The additional error in the latter bias region is attributed to the fact that the $Q$-valley ($\Gamma$-valley) in CB (VB) starts to populate with the electrons (holes) as the Fermi level is moved closer to the CB (VB) of the top (bottom) monolayer. 

\begin{figure}[t]
\centering
   \includegraphics[width=0.4\textwidth]{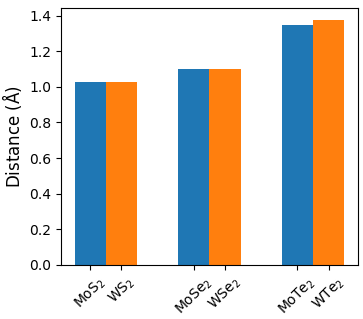}
   \caption{Tail length correction for MoS\textsubscript{2}, MoSe\textsubscript{2}, MoTe\textsubscript{2}, WS\textsubscript{2}, WSe\textsubscript{2} and WTe\textsubscript{2} determined by matching electrostatic simulations to atomistic DFT simulations.} \label{fig:IL_sep_corr}
\end{figure}

Figure figure \ref{fig:Error_Vs_bias}c (\ref{fig:Error_Vs_bias}d) shows the maximum error in CBM of the top ML (VBM of bottom ML) divided by the ML bandgap. The maximum error in CBM and VBM is presented for all TMD BL systems considered in this study with interlayer separation of 4 \AAA and 5 \AA. The bandgap normalized errors are obviously large enough ($1/4^{th}$ of the bandgap) to affect the device design quite substantially.

\begin{figure*}[t]
\centering
   \includegraphics[width=0.9\textwidth]{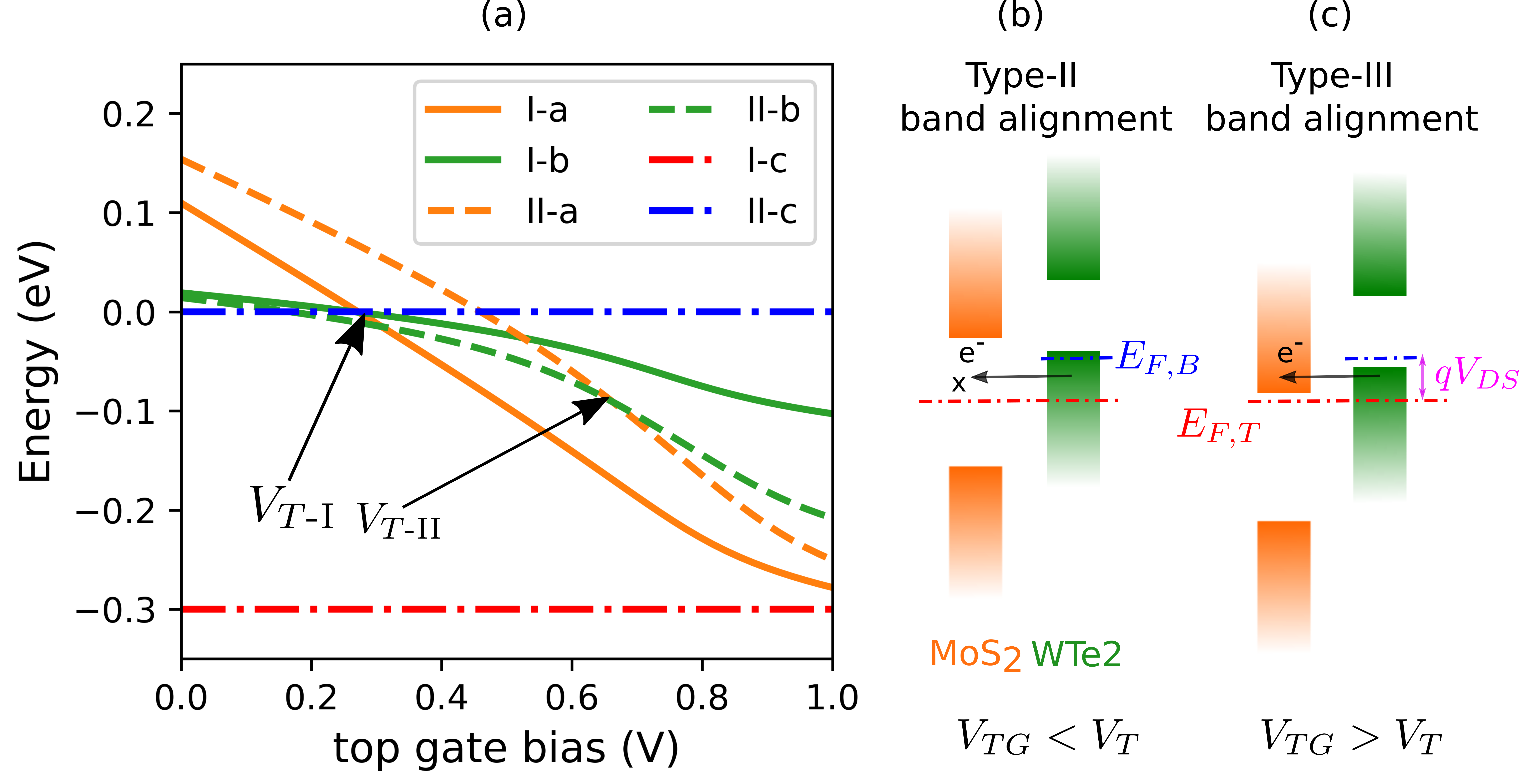}
   \caption{(a) Band alignments in MoS\textsubscript{2}-WTe\textsubscript{2} heterojunction vdW-TFET proposed by \citeauthor{Mingda_Li_JAP_2014}\cite{Mingda_Li_JAP_2014}. I-a (II-a) is CBM of top ML without (with) considering the tails of the ionic potentials. I-b (II-b) is VBM of bottom ML without (with) considering the tails of the ionic potentials. I-c (II-c) is the Fermi level of top (bottom) ML. The $V_T$ determined by electrostatic model without tail correction is $V_{T-I}=0.27 \textnormal{ V}$ where as the tail corrected electrostatic model predicts the $V_T$ to be $V_{T-II}=0.66 \textnormal{ V}$. (b) Illustration of type-II band alignment in the MoS\textsubscript{2}-WTe\textsubscript{2} heterojunction vdW-TFET for $V_{DS}=0.3$ V. The VB of WTe\textsubscript{2} is aligned in the bandgap of MoS\textsubscript{2}. And since no states are available in MoS\textsubscript{2}, for the elastic tunneling of electrons in the VB of WTe\textsubscript{2}, the BTBT current is zero. This is an off state of vdW TFET. (c) type-III band alignments in MoS\textsubscript{2}-WTe\textsubscript{2} heterojunction vdW-TFET for $V_{DS}=0.3 V$. Now plenty of states are available in the CB of MoS\textsubscript{2}, BTBT current flows and the transistor is in on state.}\label{fig:MoS2_WTe2_vdW_TFET1_Tunneling_illustration}
\end{figure*}

Since the extent of the ionic potential tails beyond the chalcogen lattice sites is comparable to the vdW gap, the electrical thickness of the thin MLs is a few angstroms more than the physical thickness determined by the lattice atom positions. Now it is verified using a finite difference 1D Poisson solver\cite{Vaidya2015} that the thickness of the MLs per se do not impact the electrostatic simulations. However,  their increased electrical thickness reduces the effective (electrical) interlayer separation. This introduces significant error in the estimation of the CBM (VBM) of the top (bottom) ML in both the non-degenerate and degenerate regions. To account for the reduction in the effective separation and achieve agreement between the electrostatic and \abinitio approaches, we introduce a simple empirical correction to the interlayer separation. The starting guess for the correction, in the illustrative case of the homogeneous BL, is twice the tail length estimated in the figure \ref{fig:Monolayer_tails}; it is then fine-tuned to match the electrostatic simulation results with that of the \abinitio. Figure \ref{fig:IL_sep_corr} presents the values of the final tail length correction for MoS\textsubscript{2}, MoSe\textsubscript{2}, MoTe\textsubscript{2}, WS\textsubscript{2}, WSe\textsubscript{2} and WTe\textsubscript{2}. Clearly, its primary dependence is on the chalcogen element in the TMD; more specifically, the atomic number thereof. This is as expected, given that potential tail in question belongs to the chalcogen atom. It may be observed from figure \ref{fig:bands_comparison} that the tail length correction leads to a near-perfect agreement between the electrostatic model and DFT simulations in the non-degenerate region. In the degenerate region, there is some discrepancy in the CBM of the top ML and VBM of the bottom ML even after applying the correction. This is more apparent from the equivalent of figure \ref{fig:bands_comparison} for the other TMDs, shown in the supplementary material. The reason behind it is the nonzero occupancy of the $Q$-valley and $\Gamma$-valley which are not included at all in the electrostatic simulations. The error for the largest bias in the CBM (VBM) of the top (bottom) ML depends inversely on the energy separation between the $K$-valley and the $Q$-valley ($\Gamma$-valley) as shown in table \ref{Table:band_parameters}. The larger the separation, the smaller would be the occupation in the higher energy valleys, and hence the smaller would be the error. The occupation of higher valleys may be included within the electrostatic model by increasing the density of states effective mass; however the detailed treatment of this correction is outside the scope of this paper.

To illustrate the impact of the tails of the ionic potential for the specific example of a vdW-TFET performance, we apply the potential tail length correction to a previously studied vdW-TFET designed using a MoS\textsubscript{2}-WTe\textsubscript{2} vertical heterojunction \cite{Mingda_Li_JAP_2014,J_Cao_TED_2016}. We use the same transistor design and simulation parameters as suggested by \citeauthor{Mingda_Li_JAP_2014}\cite{Mingda_Li_JAP_2014} and calculate the tunneling window and threshold voltage for BTBT.
The interlayer dielectric in the above design is intended to be two MLs (6 \AA) of the 2D material boron nitride (BN), which has a dielectric constant of 4.2. Their electrostatic model assumes 6 \AAA of a homogeneous material with this dielectric constant. In reality, such a MoS\textsubscript{2}-BN-WTe\textsubscript{2} vdW-TFET would be realized by stacking the MLs of MoS\textsubscript{2}, BN and WTe\textsubscript{2}, wherein each layer would be separated from its neighboring layer/s by the vdW gap. This picture suggests that while the effective dielectric constant for the BN may indeed be reasonably taken to be 4.2, the potential tail correction for the MoS\textsubscript{2} and WTe\textsubscript{2} should consider the vacuum values calculated above. The thickness of BN is then assumed to be the chalcogen-chalcogen separation minus the tail correction.
Figure \ref{fig:MoS2_WTe2_vdW_TFET1_Tunneling_illustration}a shows the CBM, VBM and Fermi levels for the MoS\textsubscript{2}-WTe\textsubscript{2} vdW-TFET when drain voltage $V_D=0.3 \textnormal{ V}$ and source is grounded, $V_S=0 \textnormal{ V}$. The gate field is varied by varying the top gate voltage while keeping the bottom gate at 0 V. It can be seen that at $V_{TG}=0\textnormal{ V}$ the band alignment is of type-II, which is illustrated in figure \ref{fig:MoS2_WTe2_vdW_TFET1_Tunneling_illustration}b. In this situation, the CBM (VBM) of the MoS\textsubscript{2} (WTe\textsubscript{2}) aligns with the bandgap of WTe\textsubscript{2} (MoS\textsubscript{2}); this is not favorable for elastic tunneling of electrons from the VB of WTe\textsubscript{2} since no states are available in the MoS\textsubscript{2}. Now, the band alignment can be modified by applying a gate field. At about $V_{TG}=0.27 \textnormal{ V}$ the band alignment changes to type-III, which is illustrated in figure \ref{fig:MoS2_WTe2_vdW_TFET1_Tunneling_illustration}c. It can be seen that the CBM of MoS\textsubscript{2} is now below the VBM of WTe\textsubscript{2} which enables BTBT. The applied bias on the top gate at which the band alignment changes from type-II to type-III may be defined as the threshold voltage $V_T$ of the vdW-TFET. The $V_T$ determined by the tail length corrected electrostatic model is 0.66 V. The pure electrostatic model underestimates the $V_T$ by 380 mV, rendering it utterly useless.

\begin{figure}[t]
\centering
   \includegraphics[width=0.45\textwidth]{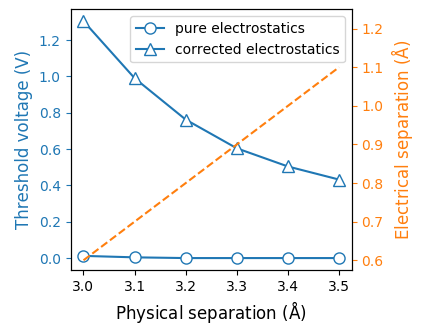}
   \caption{Threshold voltage of MoS\textsubscript{2}-WTe\textsubscript{2} heterojunction vdW-TFET calculated at $V_D=0.5 \textnormal{ V}$ and $V_S=0.0 \textnormal{ V}$ when the interlayer dielectric is vacuum. The electrostatic model predicts no variation with the interlayer separation where as the corrected electrostatic model suggests a strong dependence.} \label{fig:V_T_vs_Distnace}
\end{figure}

The situation is spectacularly aggravated when we replace BN as the interlayer dielectric by vacuum of thickness equal to vdW gap. Note that this choice of interlayer dielectric has been proposed for vdW-TFETs in the literature \cite{J_Cao_TED_2016, Mingda_Li_JEDS_2015} in order to maximize BTBT. We consider the same MoS\textsubscript{2}-WTe\textsubscript{2} BL as above, but this time with the vdW gap separation between the MLs. Since the precise value of the vdW gap for this stack was not available, we consider it to be 3-3.5 \AAA for the purpose of $V_T$ estimation. The $V_T$ estimated by the pure electrostatic model shows no variation with the interlayer separation (note that the threshold voltage coming out to be nearly 0 V in this case is coincidental - it is a function of the gate metal workfunction chosen by \citeauthor{Mingda_Li_JAP_2014}\cite{Mingda_Li_JAP_2014}). The potential tail length corrected electrostatic model, on the other hand, exhibits strong dependence as shown in figure \ref{fig:V_T_vs_Distnace}. The error in the uncorrected $V_T$ is seen to be extremely large in this scenario. This follows in turn from the much-larger relative error in the interlayer separation (viz. the difference between the physical and electrical interlayer separation, divided by the former) for the very-small vdW gap separation: note that the electrical equivalent interlayer separation is just 1.1$\textnormal{ \AA}$ (0.6$\textnormal{ \AA}$) when the physical chalcogen-chalcogen interlayer separation is 3.5$\textnormal{ \AA}$ (3$\textnormal{ \AA}$).

\section{Conclusion}
Angstrom-level insights drawn from atomistic DFT simulations suggest that pure electrostatics-based design of 2D bilayer devices may be prone to large errors. This is because the tails of the ionic potentials protrude into the interlayer gap, greatly reducing the electrical-equivalent interlayer separation with respect to the physical separation. The errors in the band energies within each ML could be as large as a fourth of the bandgap. In the example of 2D bilayer TFETs, the threshold voltage estimated from pure electrostatics may be off by hundreds of millivolts to over a volt; the smaller the physical separation, the larger the error. It is shown that employing an electrical-equivalent interlayer separation - obtained by deducting the ionic potential tail length from the physical interlayer separation - suffices to correct the electrostatic model.

\section{Methods}
\begin{figure}[b]
\centering
   \includegraphics[width=0.48\textwidth]{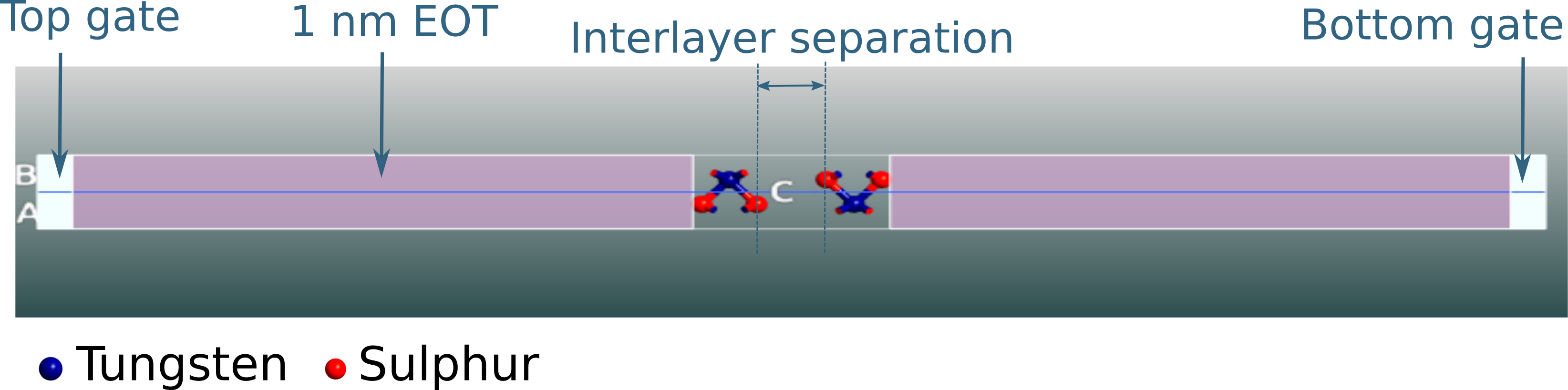}
   \caption{Simulated TMD BL system. The potential in the metallic region is the boundary condition to the Poisson's equation. The dielectric constant is set to give 1 nm EOT on both sides of the BL system.} \label{fig:simulated_slab}
\end{figure}
 For the atomistic DFT simulations here, we have used the Atomistix Simulation Toolkit (ATK) from Quantumwise \cite{ATK_Package}. Figure \ref{fig:simulated_slab} shows the simulated atomic geometry of BL WS\textsubscript{2} under an electric field. The BLs are seen to have an anti-symmetric arrangement, with each ML having a 2H geometry. We study the BL system by varying the interlayer separation between the two MLs, between 4-10 \AAA for MoS\textsubscript{2}, MoSe\textsubscript{2}, MoTe\textsubscript{2}, WS\textsubscript{2}, WSe\textsubscript{2} and WTe\textsubscript{2} BLs. The interlayer separation is defined as the distance between the lattice sites of boundary chalcogen atoms (i.e. bottom chalcogen of top ML, and top chalcogen of bottom ML). We use a 21x21x1 K-mesh for the DFT calculation, with a 110 Hartree density mesh cutoff. The convergence criterion used for the simulations is $10^{-6}$ Hartree; for the electron occupancy, we use Fermi-Dirac statistics with broadening equivalent to 300 K. We use the Open MX basis sets\cite{OMX_basis1,OMX_basis2} which are shipped along with the ATK package. We do not relax the bilayer system intentionally to focus on the interlayer separation and its impact on the band offsets. However the individual MLs are relaxed with the criteria of 0.01 eV/\AAA for force and 0.001 GPa for stress. The two relaxed MLs are then stacked to form a BL. The dielectric constant between the metallic contacts and the BL is set to give a 1 nm effective oxide thickness (EOT) i.e. 1 nm SiO\textsubscript{2}. On the other hand, the space between the two MLs is considered to be vacuum - with the dielectric constant of free space.


\bibliography{references1}

\end{document}